\documentclass[aps,twocolumn,prd,nofootinbib,showpacs]{revtex4}
\usepackage[pdftex,bookmarks]{hyperref}

\usepackage[T1]{fontenc}
\usepackage{amsmath,amssymb}

\usepackage{graphics,wrapfig,times}
\usepackage{graphicx}
\usepackage{mathrsfs}

 \makeatletter
 \newcommand{\pback}[1]{{
   \let\@rrow=\leftarrowfill
   \mathchoice{\AIN@stemPullBack{#1}{\@rrow}}{\AIN@stemPullBack{#1}{\@rrow}}
     {\AIN@indxPullBack{#1}{\@rrow}}{\AIN@indxPullBack{#1}{\@rrow}}}
   \vphantom{#1}}

 \newcommand{\AIN@stemPullBack}[2]{
   \vtop{\mathsurround=0pt
   \ialign{##\crcr$\textstyle{#1}\strut$\crcr
     \noalign{\kern-0.4ex\nointerlineskip}{\tiny#2}\crcr}}}

 \newcommand{\AIN@indxPullBack}[2]{
   \vtop{\mathsurround=0pt
   \ialign{##\crcr\hfil$\scriptstyle{#1}$\hfil\crcr
     \noalign{\kern+0.4ex\nointerlineskip}{\tiny#2}\crcr}}}

\def\be{\begin{equation}}
\def\ee{\end{equation}}
\def\bea{\begin{eqnarray}}
\def\eea{\end{eqnarray}}
\def\ba{\begin{array}}
\def\ea{\end{array}}
\def\nn{\nonumber}

\def\k{\textbf{k}}
\def\x{\textbf{x}}

\def\z{\textbf{z}}

\def\d{\textrm{d}}

\begin{document}

\title{Quantum fluctuations in planar domain-wall space-times: A possible origin of primordial preferred direction}

%

 \author{Chih-Hung Wang$^{a, c}$} \email{chwang@phy.ncu.edu.tw}
\author{Yu-Huei Wu$^{b, c}$} \email{yhwu@phy.ncu.edu.tw}
\author{Stephen~D.~H.~Hsu$^{d}$} \email{hsu@uoregon.edu}
\affiliation{$^{a}$ Department of Physics, Tamkang University, Taipei 25137, Taiwan, R.O.C. \\
$^{b}$ Center for Mathematics and Theoretical Physics, National Central University, Chungli 320, Taiwan, R.O.C.\\
$^{c}$ Department of Physics, National Central University, Chungli 320, Taiwan, R.O.C.\\
$^{d}$Institute of Theoretical Science, University of Oregon,
Eugene, OR 97403, USA}





\begin{abstract}

We study the gravitational effects of a planar domain wall on quantum fluctuations of a massless scalar field during inflation. By obtaining an exact solution of the scalar field equation in de Sitter space, we show that the gravitational effects of the domain wall break the rotational invariance of the primordial power spectrum without affecting the translational invariance. The strength of rotational violation is determined by one dimensionless parameter $\beta$, which is a function of two physical parameters, the domain wall surface tension $\sigma$ and cosmological constant $\Lambda$. In the limit of small $\beta$, the leading effect of rotational violation of the primordial power spectrum is scale-invariant.
\end{abstract}
\date{\today}
\maketitle


\emph{Introduction.}~Inflationary cosmology was originally proposed to solve the horizon, flatness, and monopole problems \cite{Guth-81, LAS-82, L-90}. The monopole problem, or, more generally, the topological defect problem, arises when an early epoch of symmetry breaking produces defects such as mononpoles, cosmic strings or domain walls. These objects redshift more slowly than radiation and would come to dominate the energy density of the Universe (or leave a signal in the CMB), in conflict with observation \cite{ZV}.  Because physical distances increase exponentially during inflation, the number density of topological defects is driven to zero. Heavy topological defects (i.e., associated with energy scales larger than that of inflation) then presumably leave almost no detectable evidence, and the only defects we can observe today are those arising from phase transitions which occurred after inflation \cite{SV}.  The formation and evolution of domain walls during inflation and background Friedmann-Robertson-Walker (FRW) Universe has been studied analytically in \cite{BBGH-95}.

However, the gravitational fields of domain walls can affect the primordial quantum fluctuations of scalar fields during inflation. Quantum fluctuations of scalar fields in de Sitter space-time have been widely studied (see \cite{BD-82} and the references in) and lead to the well-known scale-invariant or Harrison-Zel'dovich spectrum \cite{BST-83, L-90, LL-00}.  In this Letter, we study quantum fluctuations of a massless scalar field $\phi$ in planar domain-wall space-times with positive $\Lambda$, where stationary observers feel a repulsive constant gravitational force normal to the wall. The gravitational fields of a planar domain wall will break the rotational symmetry of the primordial power spectrum, so the occupation numbers of modes $\k_n$ perpendicular to the wall are different from those of modes $\k_{\|}$ parallel to it.



\emph{The basic picture}~ Monte Carlo simulations of domain wall formation indicate that a typical system of domain walls is dominated by an infinite wall in addition to some closed walls, most having radius of curvature $R \sim \xi$, where the correlation length $\xi$ is smaller than the horizon size $1/H$ due to causality \cite{VS}. Moreover, simulations of domain wall evolution in radiation and matter dominated Universes indicate that each horizon typically contains one large domain wall, which extends across the horizon \cite{PRS-89}. We consider a large domain wall with radius $R$ several times larger than a pre-inflationary horizon volume, and stationary with respect to cosmic time evolution.

To completely describe the resulting gravitational fields in a pre-inflationary horizon, which contains a portion of the large wall, we must solve the Einstein field equations taking into account the wall. However, since our present horizon $1/H_0$ was originally a small part of the pre-inflationary horizon, it is sufficient to study the gravitational fields in a small patch. For any point $p$ on a closed wall, one can define a local neighborhood $\mathcal{N}_p$, which is a sphere with radius $r$ satisfying $r << R$ and center at $p$. In $\mathcal{N}_p$, the large (but ultimately closed) wall can be well approximated as an infinite planar wall. The dynamics of spherical domain walls in de Sitter space has been studied in \cite{bubble}. Using their results, it can be shown that large bubbles of the type we consider have physical radius $R$ which increases exponentially during inflation: $R \sim R_0 \exp ({\sqrt{\Lambda / 3}\tau})$. This implies that even as the physical distance from the precursor of our visible universe to the domain wall grows during inflation, so do the size and radius of the wall, {\it preserving the approximation of an infinite planar wall}.

In other words, if our visible Universe is well inside $\mathcal{N}_p$ at the end of inflation (i.e., the final inflated size $r_f$ is larger than $1/H_0$), we expect to see gravitational effects which approximate those of an infinite planar wall. It is not difficult to satisfy the condition $r_f> 1/H_0$ by requiring a sufficient number of $e$-foldings. So, although inflation ensures that there are no domain walls in our present horizon volume after inflation, the gravitational effect of the large wall whose existence we assume at the beginning of inflation can still affect primordial quantum fluctuations {\it during} inflation, which leave an imprint on the CMB. The initial conditions assumed in this basic picture are plausible (i.e., not highly improbable), given the main assumption that domain walls formed before inflation.

%
The gravitational effects of domain walls in the thin-wall approximation have been discussed in \cite{ISV, CGS-93, WCW-11}. In particularly,  \cite{WCW-11} found an exact solution of an infinite planar domain wall space-time with positive cosmological constant  $\Lambda$ (see eq. (\ref{dw-1})) in the co-moving coordinates, where the wall is sitting at $z=0$, and when surface tension $\sigma$ of the domain wall vanishes, the metric (\ref{dw-1}) returns to the steady-state Universe in conformal time (see eq. (\ref{dw-2})).  So one can define the vacuum state for both $z>0$ and $z<0$ sides in the background metric (\ref{dw-1}) by assuming that when $\sigma$ vanishes, the vacuum state will correspond to the Bunch-Davies vacuum \cite{BD-87}.  With this definition of the vacuum state, we study the gravitational effects of the planar domain wall on the quantum fluctuations of a massless scalar field.
We then find that rotational invariance of all of the fluctuation modes is broken by the gravitational fields of the domain wall and the resulting rotational violation in the primordial power spectrum turns out to be scale-invariant.

It is known that the physical wavelengths of fluctuation modes increase exponentially during inflation: $\lambda_p \sim \lambda \exp ({\sqrt{\Lambda / 3 }\tau})$. Once the fluctuation modes cross the horizon, they become classical and their values become constant \cite{LL-00}. By associating these fluctuations with curvature perturbations, which are constant well outside the horizon, we obtain initial values for density perturbations. Large-scale modes, which remain outside the horizon at the last-scattering surface, will imprint their primordial values on CMB anisotropies. Since the power spectrum is calculated in co-moving coordinates, where the planar wall sits at $z=0$,  it is necessary to translate the original point to our galaxy, which is the center of our present horizon in CMB observations. Because the power spectrum is translation invariant, this coordinate translation does not cause any change in our results.

\emph{Planar domain walls in de-Sitter space-time.}~  In the standard cosmological model, the Universe is homogeneous and isotropic with respect to cosmic time evolution. Domain walls, once formed, will evolve to minimize their surface area, subject to interactions with the background enviroment \cite{VS}. If the interactions are significant this motion can be overdamped and the wall motion relatively slow. We neglect any motion relative to the thermal rest frame and take the wall to be co-moving along the cosmic time direction.

The metric of a planar domain wall in de-Sitter space-time with reflection symmetry has been obtained in \cite{WCW-11}:
\bea
d s^2 = \frac{1}{\alpha^2 \left(\eta + \beta |z|\right)^2} (- d \eta^2 + d z^2 + d x^2 + d y^2), \label{dw-1}
\eea where the wall is placed at $z=0$. $\alpha= \sqrt{\Lambda / 12 \Gamma} (\Gamma+1)$,  $\beta= (\Gamma-1)/(\Gamma+1)$, satisfying $-1<\beta\leqslant0$, and  $\Gamma$ is a dimensionless parameter
\bea
 \Gamma= 1+\frac{3\epsilon- \sqrt{48 \epsilon + 9 \epsilon^2}}{8}, \label{gamma}
 \eea
where $\epsilon=\kappa^2\sigma^2/\Lambda$ and $\sigma$ is the surface tension of the domain wall.  Eq. (\ref{gamma}), which gives $0<\Gamma\leqslant 1$, is only valid for the coordinate ranges $-\infty <\eta + \beta|z|<0$. When $\sigma=0$, the metric (\ref{dw-1}) is simply that of a steady-state Universe in the conformal time \cite{BD-82}.  Throughout this Letter we use the units $c=\hbar=1$ and $\kappa=8\pi G$.

Domain walls produce repulsive  gravitational forces \cite{ISV}.  To understand the gravitational effects of metric (\ref{dw-1}), we consider observers stationary relative to the wall on the $z>0$ side, with 4-velocities described by a future-pointing unit time-like vector field $U=-\alpha(\eta+ \beta z)\partial_\eta$. Their 4-acceleration, which is defined by $\mathcal{A} \equiv \nabla_{U} {U}$, has a constant magnitude $|\mathcal{A}|\equiv\sqrt{g(\mathcal{A}, \mathcal{A})}=|\alpha \beta|=  \kappa \sigma / 4$ and $z$-component $A_z\equiv g(\nabla_{U} {U}, - \alpha(\eta + \beta z) \partial_z)=- \kappa \sigma / 4$, where the minus sign denotes the acceleration toward the wall. This implies that the gravitational field of a planar domain wall produces a constant repulsive force on each observer, independent of their distance from the wall. For this reason, translation invariance is not violated by the gravitational field of the wall. Because of the reflection symmetry, the same is true for the $z<0$ side.

In the coordinates
%
$\check{\eta}= (\eta + \beta z ) /  \sqrt{1-\beta^2}$,  $\check{z}= (z + \beta \eta ) /  \sqrt{1-\beta^2}$, $\check{x}=x$ and $\check{y}=y$,
 %
the metric (\ref{dw-1}) becomes (for $z > 0$)
 \bea
d s^2 = \frac{1}{\frac{\Lambda}{3}\check{\eta}^2}(- d \check{\eta}^2 + d \check{z}^2 +d \check{x} ^2 + d \check{y}^2), \label{dw-2}
\eea  which describes a steady-state Universe in conformal time. However, if one uses the coordinates ($\check{\eta}, \check{z}, \check{x}, \check{y}$) on the $z<0$ side,  the metric becomes $d s^2 = ( \frac{\Lambda}{3} (\check{\eta}- \frac{2\beta \check{z}}{1-\beta^2})^2 )^{-1}
(- d \check{\eta}^2 + d \check{z}^2 +d \check{x} ^2 + d \check{y}^2)$.
%
The coordinate transformations between ($\eta, z, x, y$) and ($\check{\eta}, \check{z}, \check{x}, \check{y}$)
are very similar to Lorentz transformations (boosts) and $\beta$ is analogous to the relative velocity of two inertial frames. One may notice that the wall is not stationary in ($\check{\eta}, \check{z}, \check{x}, \check{y}$).  It is known that the motion of an uniformly accelerated observer $O$ along the $x^1$ direction in Minkowski space-time with the Minkowski coordinates $(x^0, x^1, x^2, x^3)$ is described by $x^0=A^{-1}\sinh{A\tau}, x^1=A^{-1}\cosh{A\tau}$ and $(x^2, x^3)=\textrm{const.}$, where $A=|\mathcal{A}|$ and $\tau$ is the proper time of the observer \cite{MTW}. So $O$'s trajectory is hyperbolic, i.e. $(x^1)^2 - (x^0)^2 = A^{-2}$, in Minkowski space-time.
However, the wall's motion in de-Sitter space with the coordinates ($\check{\eta}, \check{z}, \check{x}, \check{y}$) gives
$\check{\eta}=  - e^{-\alpha\tau}  / \alpha \sqrt{1-\beta^2}$ and $\check{z}=-\beta e^{-\alpha\tau}  / \alpha \sqrt{1-\beta^2}$, where $\tau$ is the wall's proper time. It turns out that the wall's trajectory, which has constant magnitude of acceleration $|\alpha\beta|$, is a straight line $\check{z}= \beta \,\check{\eta}$ in de-Sitter space with the coordinates ($\check{\eta}, \check{z}, \check{x}, \check{y}$).
In the coordinates ($\check{\eta}, \check{z}, \check{x}, \check{y}$), stationary observers, who are in relative motion with respect to the wall and  4-velocities $-\sqrt{\Lambda / 3}\check{\eta}\,\partial_{\check{\eta}}$, will follow geodesics. We conclude that the stationary observers associated with two different coordinates ($\eta, z, x, y$) and ($\check{\eta}, \check{z}, \check{x}, \check{y}$) will correspond to uniformly accelerated observers and geodesic observers, respectively.


Before we discuss quantum fluctuations, it is helpful to describe the metric (\ref{dw-1}) by introducing proper-time coordinate $\tau=- \frac{1}{\alpha}\ln[-\alpha (\eta \pm \beta z ) ]$ and $z'=\sqrt{1-\beta^2} z$, so Eq. (\ref{dw-1}) becomes
\bea
d s^2 = - d \tau^2 \pm  \frac{2 \beta\, e^{\alpha \tau} }{\sqrt{1-\beta^2}} d \tau d z' + e^{2 \alpha \tau}(d z'^2 + d x^2 + d y^2), \label{dw-3}
\eea where $\pm$ corresponds to $z'>0$ and $z'<0$ sides, respectively. It is clear that the metric (\ref{dw-3}) also has reflection symmetry about $z'=0$.  Moreover, the stationary observers, whose 4-velocities are $\partial_\tau$, also have constant acceleration $|\mathcal{A}|=|\alpha\beta|$. For $\beta=0$, i.e. $\sigma=0$, the metric (\ref{dw-3}) becomes the metric (\ref{dw-2}) in ($\check{\tau}, \check{z}, \check{x}, \check{y}$) coordinates, where $\check{\tau}=-\sqrt{3 / \Lambda}\ln(-\sqrt{\Lambda / 3}\check{\eta})$.  Since the metric (\ref{dw-1}) is $z$-dependent, one might expect that the primordial density fluctuations will violate translational invariance. On the other hand, as discussed above, the gravitational force due to a planar domain wall is $z$-independent, so density fluctuations should be translationally invariant.  From the metric (\ref{dw-3}),  it becomes clear that the primordial power spectrum will be translation invariant, since the metric (\ref{dw-3}) only depends on $\tau$. Moreover, the appearance of the cross term $g_{\tau z}$ indicates that the gravitational effects of planar domain walls will break the rotational invariance, i.e.  $O(3)$ symmetry, of the power spectrum. In the post-Newtonian theory \cite{MTW}, the metric components $g_{0i}$ are associated with the 3-velocities of gravitating sources.


\emph{Quantum fluctuations in planar domain-wall space-times.}
Quantum fluctuations in de-Sitter space-time have been widely studied \cite{L-90, BD-82, BD-87}. In particular, it is known that time-like geodesic observers in de-Sitter space-time will detect thermal radiation with temperature $T=\sqrt{\Lambda / 12 \pi^2}$ \cite{GH-77}. A stationary observer with 4-velocity $\partial_\tau$ in de-Sitter space-time, which is described by the metric (\ref{dw-3}) with vanishing $\beta$, will perceive an isotropic thermal bath of radiation \cite{BD-82}. However, in the presence of a planar domain wall the stationary observer with velocity $\partial_\tau$ has constant acceleration $A_z$, so one should expect that, besides the particle production due to the de-Sitter horizon, the constantly accelerating observer should detect extra particles, which are associated with the acceleration $A_z$.  A well-known example is the Unruh effect, which shows that an observer constantly accelerating along the $z$-axis in Minkowski space-time will see particles with temperature $T= A_z / 2\pi$, though an inertial observer will detect no particles \cite{U-76}.

To understand the gravitational effects of a planar domain wall on primordial density fluctuations, we start from a massless scalar field $\phi$ satisfying the field equation
\bea
\d\star \d \phi =0, \label{sc-1}
\eea where $\d$ is the exterior derivative and $\star$ is the Hodge map associated with the metric $g$. Mode functions $\phi_{\check{\k}}$, which are exact solutions for $z>0$, are
\bea
\phi_{\check{\k}}(\check{x}^i)=\check{\eta}^{\frac{3}{2}}\left[c_1(\check{k}) H^{(1)}_{3/2}(\check{k}\check{\eta})+ c_2(\check{k}) \,H^{(2)}_{3/2}(\check{k}\check{\eta})\right] e^{i \check{\k}\cdot \check{\x}}, \label{sc-2}
\eea where $H^{(i)}_{3/2}$ are Hankel functions, $\check{k}\equiv ( \check{k}_z^2 + \check{k}_x^2+ \check{k}_y^2 )^{1/2}$ and $\check{x}^i=(\check{\eta},\, \check{\x})$.
For simplicity, we will only consider the solution $\phi_{\check{\k}}$ for $z>0$. Using reflection symmetry, the $z<0$ solution can be obtained. By noting that the metric (\ref{dw-1}) in the coordinates ($\check{\eta}, \check{z}, \check{x}, \check{y}$) is the metric (\ref{dw-2}), the normalization of Eq. (\ref{sc-2}) is straightforward and gives $|c_2(\check{k})|^2- |c_1(\check{k})|^2=\pi \Lambda / 12$. The choice of $c_1(\check{k})$ and $c_2(\check{k})$ corresponds to the choice of vacuum state \cite{BD-87}. We require that when $\beta=0$, the vacuum state is identical to the Bunch-Davies vacuum, i.e. $c_1(\check{k})=0$ and $c_2(\check{k}) =\sqrt{\pi \Lambda / 12}$ \cite{L-90, BD-87}.

Since $\check{z}$ depends on the variable $\eta$,  we should rewrite Eq. (\ref{sc-2}) in the coordinates $(\eta, z, x, y)$. Furthermore, it is useful to introduce the proper-time $\tau$, which satisfies $e^{-\alpha\tau}=-\alpha(\eta + \beta z)$, so Eq. (\ref{sc-2}) becomes
\begin{widetext}
\bea
\phi_{\k}(x^i) = \sqrt{\frac{\Lambda}{6}} \frac{1}{k\sqrt{k}}\left(\frac{1+ \beta \hat{\k}\cdot \hat{\z}}{\sqrt{1-\beta^2}}\right)^{-3/2}\left(i + \frac{k }{\alpha(1-\beta^2)}e^{-\alpha\tau}(1+ \beta \hat{\k}\cdot \hat{\z}) \right) e^{  i (k_z+ \beta k) z + i k_x x +  i k_y y + i \frac{k}{\alpha}e^{-\alpha \tau} },\label{sc-3}
%
\eea
\end{widetext} where $x^i=(\tau, x, y, z)$, $k\equiv ( k_z^2+k_x^2 + k_y^2 )^{1/2}= \frac{\check{k}- \beta \check{k}_z}{\sqrt{1-\beta^2}}$, $k_z=\frac{\check{k}_z- \beta \check{k}}{\sqrt{1-\beta^2}}$,
$k_x=\check{k}_x$ and $k_{y}= \check{k}_y$. $\hat{\k}$, $\hat{\z}$ are unit vectors and $\hat{\k}\cdot \hat{\z}=k_z / k$. When $\beta$ goes to zero, Eq. (\ref{sc-3}) returns to the well-known solution of a massless scalar field in de-Sitter space-time.   Moreover, the $\beta \hat{\k}\cdot \hat{\z}$ terms indicate the existence of a preferred direction in the primordial density spectrum. Eq. (\ref{sc-3}) also tells us that the rotational violation will appear not only in the low-frequency $k$ modes but also high-frequency modes. The gravitational effects of the constant acceleration will affect all frequency modes.

To quantize the $\phi$ field, one may expand $\phi$ in creation and annihilation operators, $a_{\k}{\dagger}$ and $a_{\k}$, as
\bea
\phi= \int \frac{d^3 k}{(2\pi)^{3/2}} ~~  a_{\k}\phi_{\k}(x^i) + a_{\k}^{\dagger} \phi_{\k}^{*}(x^i) ,
\eea
with the vacuum state $|0\rangle$, satisfying $a_{\k} |0\rangle=0$. The vacuum expectation value of $\phi^2$ is
$
\langle \phi^2(x^i) \rangle=\frac{1}{(2\pi)^3}\int|\phi_{\k}(x^i)|^2 d^3 k.
$ It is convenient to introduce physical momenta $p=k e^{-\alpha\tau}$, which are exponentially decreasing during inflation, to obtain
\bea
&&\langle 0 \vert \phi^2(x^i) \vert 0  \rangle= \nn\\
&&\int \frac{d^3 p}{(2 \pi p)^3} \left[ \frac{\Lambda}{6} \left(\frac{1\pm \beta \hat{\k}\cdot \hat{\z}}{\sqrt{1-\beta^2}}\right)^{-3}+ \frac{\sqrt{1-\beta^2}\, p^2}{ 2(1\pm \beta \hat{\k}\cdot \hat{\z})}\right]  \label{sc-3-2}
\eea where $\pm$ denotes $\langle \phi^2(x^i) \rangle$ for $z>0$ and $z<0$ sides, respectively. For those $p$ modes with  physical wavelengths $\lambda_p$ well inside the horizon, i.e. $p\gg \sqrt{\Lambda / 3}$, the second term of Eq. (\ref{sc-3-2}) dominates and $\beta=0$ simply gives the vacuum fluctuations in Minkowski space-time: $\frac{1}{(2\pi)^3}\int \frac{d^3 p}{2 p}$. However, when $\lambda_p$ crosses the horizon, i.e. $p\lesssim  \sqrt{\Lambda / 3}$, the first term, which is time-independent, becomes dominant and taking $\beta=0$ yields the well-known scale-invariant Harrison-Zeldovich spectrum.

For the fluctuation modes which are well outside the horizon at $\tau=\tau_*$ during inflation, the spectrum of the scalar field fluctuations ${P}_\phi (\k, \tau_*) =  |\phi_\k(\tau_*)|^2$ gives
\bea
{P}_\phi=\frac{\Lambda (1-\beta^2)^{\frac{3}{2}}}{12\,k^3} \left[  ({1+ \beta \hat{\textbf{k}}\cdot \hat{\textbf{z}}})^{-3} + ({1- \beta \hat{\textbf{k}}\cdot \hat{\textbf{z}}})^{-3} \right], \label{sc-3-3}
\eea where ${P}_\phi (\k)$ has been made to satisfy reflection symmetry, i.e. ${P}_\phi(\k) = {P}_\phi(-\k)$. So Eq. (\ref{sc-3-3}) is valid for both $z>0$ and $z<0$.

We should now discuss how Eq. (\ref{sc-3-3}) relates to primordial curvature perturbations $\mathcal{R}_\k$, which are the initial values for density perturbations $\delta_\k$, and have constant value outside the horizon.  Since the evolution of perturbed classical quantities are described in a background homogenous and isotropic Universe, we should study those quantities in the geodesic coordinates ($\check{\eta}, \check{z}, \check{x}, \check{y}$), where the metric is described by Eq. (\ref{dw-2}) for the $z>0$ side. Taking our observable Universe to be located at $z>0$, we transform Eq. (\ref{sc-3-3}), which is the power spectrum for classical field configurations, to the coordinates ($\check{\eta}, \check{z}, \check{x}, \check{y}$). It becomes
 \bea
 {P}_\phi({\check{\textbf{k}}})=\frac{\Lambda}{12} \left[ \check{k}^{-3} + \big(\check{k}-  \frac{2\,\beta}{1-\beta^2}\,\check{\textbf{k}}\cdot \hat{\textbf{z}}\big)^{-3} \right], \label{sc-3-4}
 \eea which is only valid for $z>0$.
The primordial power spectrum of curvature perturbation is equal to $\mathcal{P}_{\mathcal{R}}(\check{\k})=(\check{k}^3/2\pi^2) [(H/\dot{\varphi})^2 P_\phi(\check{\k}, \tau)]_{\tau=\tau_*}$, where $\varphi(\tau)$ denotes the background slow-roll inflaton field. In the slow-roll inflation, we obtain
\bea
\mathcal{P}_{\mathcal{R}}(\check{\k}) = \frac{V(\varphi)}{48\pi^2 M_{\textrm{Pl}}^4\,\varepsilon} \left[ 1 + \big(\,1-  \frac{2\,\beta}{1-\beta^2}\, \frac{\check{\textbf{k}}\cdot \hat{\textbf{z}}}{\check{k}}\,\big)^{-3} \right], \label{sc-3-5}
\eea where $V(\varphi)$ is a slow-roll potential and $\varepsilon= \frac{1}{2} M_{\textrm{Pl}}^2 (V'/V)$ is one of the slow-roll parameters \cite{LL-00}.
%
In order to associate $\mathcal{P}_{\mathcal{R}}(\check{\k})$ to CMB anisotropies, one should translate the original point to the location of our galaxy. However, because of the translational invariance of  $\mathcal{P}_{\mathcal{R}}(\check{\k})$, such translation does not change the results.

The requirement $|\beta| \ll 1$ yields a constraint on the surface tension of the domain wall: $\sigma \ll M_{\textrm{Pl}} V^{1/2}$. In this limit the leading-order effect of rotational violation of $\mathcal{P}_{\mathcal{R}}(\check{\k})$ is
\bea
\mathcal{P}_{\mathcal{R}}(\check{\k}) = \frac{V(\varphi)}{24\pi^2 M_{\textrm{Pl}}^4\,\varepsilon} \left[ 1 + 3 \beta\, \frac{\check{\textbf{k}}\cdot \hat{\textbf{z}}}{\check{k}} +\cdots \right], \label{sc-3-6}
\eea which may correspond to a primordial dipole \cite{HL-09}.
 Eq. (\ref{sc-3-5}) indicates that magnitudes of rotational symmetry violation do not depend on frequency $k$, so the rotational violation of $\mathcal{P}_{\mathcal{R}}(\check{\k})$ is scale-invarant.  Of course, our results only apply to the local neighborhood $\mathcal{N}_p$, where one can approximate a realistic domain wall
by a planar infinite wall.  If our observable Universe is not well inside the $\mathcal{N}_p$ defined at the end of inflation, one should expect to obtain not only rotational but also translational violation of primordial power spectrum, due to the curvature (deviation from planarity) of the wall.

\emph{Discussion.}~
Because discrete symmetries are common in models of fundamental physics, domain walls are a particularly plausible type of topological defect. However, their effects are so strong that one can largely rule them out in the post-inflationary big bang. (See \cite{Abbott-09} for discussion of how to detect domain walls using gravitational waves.) In this Letter we found analytical solutions of the scalar field equation in the gravitational background of a planar domain wall and positive $\Lambda$. Moreover, the planar domain-wall metric (\ref{dw-1}) in the co-moving coordinates ($\eta, z, x, y$), which cover both $z>0$ and $z<0$ sides, allows us to define the vacuum state by assuming that when $\sigma$ vanishes, the vacuum state should correspond to the Bunch-Davies vacuum. The primordial power spectrum of the scalar field $P_\phi(\k)$, which satisfies $P_\phi(\k)=P_\phi(-\k),$ does not have rotational invariance. We further calculate the power spectrum of primordial curvature perturbations $\mathcal{P}_{\mathcal{R}}(\check{\k}) $, which is directly related to primordial density perturbations on superhorizon scales.

Our main conclusion is that even as domain walls are inflated away they leave a characteristic imprint on the quantum fluctuations of the inflaton, which could possibly lead to observable CMB anisotropies. Under the approximation $|\beta| \ll 1$, the leading effect of rotational violation corresponds to  a primordial dipole in $\mathcal{P}_{\mathcal{R}}(\check{\k})$.

Violation of rotational and translational symmetry in the primordial power spectrum has been investigated recently \cite{ACW-07}, motivated in part by possible large-scale CMB anomalies, e.g.  alignments of low multi-poles in CMB anisotropies \cite{TOH-03, SSHC-04}. Furthermore, the possible existence of CMB statistical anisotropy has also been examined (for example, see \cite{HL-09, MEC-11}). Our future work will investigate the correlations of multipole moments of CMB anisotropies $\langle a_{lm} a^*{_{l'm'}}\rangle$ implied by Eq. (\ref{sc-3-5}), to deduce constraints on the parameter $\beta$ from observational data.



\emph{Acknowledgement}~
CHW and YHW are thankful for interesting discussion with  Prof Hing-Tong Cho and Dr Jen-Tsung Hsiang. YHW would like to thank Prof James M. Nester and Prof. Peilong Chen for their encouragement.  CHW is supported by the National Science Council of the Republic of China under the grants NSC 98-2811-M-032-005 and YHW is fully supported by the NCU Top University Project funded by the Ministry of Education, Taiwan ROC. SH thanks Academia Sinica for its hospitality while this work was initiated, and acknowledges support from the US Department of Energy (Grant No. DE-FG02-96ER40949) and the National Science Council of Taiwan.

%


\end{document}